\documentclass[conference]{IEEEtran}
\IEEEoverridecommandlockouts
\usepackage{cite}
\usepackage{amsmath,amssymb,amsfonts}
\usepackage{algorithmic}
\usepackage{graphicx}
\usepackage{textcomp}
\usepackage{xcolor}
 \usepackage{adjustbox}
 \usepackage{booktabs}
 \usepackage{hyperref}
\def\BibTeX{{\rm B\kern-.05em{\sc i\kern-.025em b}\kern-.08em
    T\kern-.1667em\lower.7ex\hbox{E}\kern-.125emX}}

\definecolor{BoxGray}{gray}{0.93}

\definecolor{BoxGray}{gray}{0.93}
\long\def\finding#1{\hspace{-0.5cm} \colorbox{BoxGray}{\fbox{\parbox{0.95\columnwidth}{\emph{#1}}}}}

\newif\ifrebuttal
\rebuttaltrue

\newcounter{reviewer}
\newcounter{comment}[reviewer]
\newcounter{shepherd}
\newcounter{scomment}[shepherd]
\renewcommand{\thereviewer}{\Alph{reviewer}}
\renewcommand{\thecomment}{\thereviewer.\arabic{comment}}
\newcommand{\thesheperd}{X\Alph{shepherd}}
\renewcommand{\thescomment}{\thesheperd.\arabic{scomment}}
\def\reviewer{%
	\refstepcounter{reviewer}%
	\vspace*{2\baselineskip}\section*{\textsf{Comments from Reviewer } \thereviewer}\vspace*{-0.5\baselineskip}}
\def\comment{%
	\refstepcounter{comment}%
	\par\vspace*{\baselineskip}\noindent{\textbf{Comment\,\thecomment)}} \ignorespaces\tt}
\def\shepherd{%
	\refstepcounter{shepherd}%
	\section*{\textsf{Comments Common to All Reviewers}\vspace*{-\baselineskip}}}
\def\scomment{%
	\refstepcounter{scomment}%
	\par\vspace*{\baselineskip}\noindent{\textbf{Comment\,\thescomment)}} \ignorespaces\sc}
\def\reply{\rm\vspace*{0.5\baselineskip}\par\noindent{\textbf{Reply }}}

\let\mylabel\label

\long\def\fixed#1{\ifrebuttal {\color{red}#1}\else{#1}\fi}

\def\reviewnote#1{\ifrebuttal\expandafter\mylabel{back:#1}\marginpar[\ref{#1}]{\ref{#1}}\fi}
\def\reviewnotemulti#1{\ifrebuttal\expandafter\marginpar{#1}\fi}

\begin{document}

\rebuttalfalse

\ifrebuttal
    \onecolumn
    \pagestyle{plain}
    \renewcommand{\thepage}{\roman{page}}
    \renewcommand{\thetable}{\arabic{table} for Reviewer \thereviewer}


\shepherd
\scomment \label{X:Notation}
Notation for comments

\reply Each comment is marked as M.n to mark the n-th comment of reviewer M. They are further tracked in the text, while added text is marked \fixed{in a different color}. Small edits of text (e.g., shortening a paragraph to eliminate an orphan) are not marked.

\scomment \label{X:TOST}
Statistical Equivalence vs No Statistical Difference

\reply We used a statistical test for \emph{significant equivalence} which is not the same as
\emph{No significant difference}. Test for equivalences are very common in drug research and very rare in software engineering.
This might be the source of confusion for some of comments (e.g., \ref{A:elaborate:on:equivalence}, \ref{C:null:and:alternative}, \ref{C:multiple:tests}). 

We used TOST (Two One-Sided Tests) as a test of equivalence which was initially proposed by Schuirmann et al.~\cite{schuirmann1981hypothesis} and is widely used in pharmacological and food sciences to check whether the two treatments are equivalent within a specified range (either as an additive constant or as a ratio)~\cite{food2001guidance, meyners2012equivalence}.
For the US Food and Drug Administration and the European Medicine Agency, two drugs are considered equivalent if their respective distributions $x$ and $y$ are such that $x \cdot \rho < y$ (one-sided test) and  $y < x \cdot 1/\rho$ (the second one-sided test) for $\rho = 0.8$. The results of the two tests are conservatively combined by taking the maximum of the two $p$-values. 

The null hypothesis of a TOST is no equivalence between two samples and the alternative hypothesis is that there is an equivalence. In the  null hypothesis one performances might be in part much lower than the other (but not uniformly lower to have a statistical difference in one direction) or too high or too spread to be comparable.
 
The underlying directional test can be chosen depending on the conditions at hand. In our study, we used Mann–Whitney U test as the underlying test because we cannot assume normality (see also \ref{C:test:assumptions}).

We added the explanation here and put a terser explanation with citations in the body (see \ref{X:examples} for an overall justification of this and similar choices). 

\scomment \label{X:examples} Explanations vs Citations
\reply Unfortunately, the rebuttal keeps the size of the paper constant (6\, pages+1) while giving us 6 pages for the rebuttal. While we feel that 4\, rebuttal + 8\, paper +1\,  references we would have allowed us to incorporate everything in the main text we have to work with the current rules. So we provided more information here and are terser in the main body (e.g. added citations to reference sources for \ref{X:TOST} and \ref{A:manual:analysis}) rather than formulas and text). We will add explanations to the full paper.

\reviewer 

\comment \label{A:elaborate:on:equivalence}
\label{A:elaborate:on:equivalence:second}
I think the overall idea and purpose of the paper to find out if LLM's success stems from memorization is both useful and relevant.
The paper hypothesizes that any deviation in the exact location of the error should be enough to show if an LLM relies on memorization or not. It then states that no significant difference in performance is expected between small and large displacements without elaborating further.

\reply We thank the reviewer for appreciating our work.
We have rephrased the relevant section, introducing the notions of the ``cheater" and the ``ideal" LLM and how a bayesian approach can help disciminating the result (see page \pageref{back:A:elaborate:on:equivalence}).

We explained that if LLMs truly understand code and reason about vulnerabilities (an ideal LLMs) any displacement in the vulnerability location would result in a highly unlike event: a successful patch repair (a code change), as we are suggesting the model to change a line that should not be changed. If instead LLMs just memorize patterns, their behavior should remain largely unaffected by localization errors, provided they can still recognize the code they saw. A similar idea was submitted in parallel to our submission to FSE'25 for repair of normal defects and we added it here \cite{kong2025demystifying}. Their result is less strong than our potential result as they did not employ the test for equivalence.

\comment \label{A:displacements} The assumption that no significant difference in performance would be observed regardless if a line is moved very close or very far to the original line seems questionable. Especially since the authors purposefully choose a fairly small range (1-4 lines displacement) for their analysis. I recommend the authors to elaborate and justify more in the paper why this assumption and this methodology are sound and should lead to trustworthy results.

\reply Please refer to \ref{X:TOST} as we don't look for 'no significant difference' but for 'significant equivalence' (see also \ref{B:research:questions}). We revised our paper to follow this suggestion by expanding the justifications, and by using larger displacements to make the hypothesis stronger. A windows of 3 lines is the git diff size and this might have been seen by the LLM, which might have learned it. The same lenght to consider a displacement correct has been used by human experiments \cite{papotti2024effects}.
So we proposed displacement is now $\pm2~\pm4~\pm8$. In  Table, we report how many entries in VJBench can support each displacement without the marked line falling outside of the vulnerable function. This is explained in the paper at page \pageref{back:A:displacements}.

\begin{table}[h]
\centering
\caption{Number of Vulnerabilies where displacement would still fit in the method} \label{tab:displacement}
\begin{tabular}{crrrr}
 \hline
 & \multicolumn{4}{c}{Error displacement}  \\
sign & 0 & 2 & 4 & 8 \\ \hline
+ & 50 & 50 & 46 & 27 \\ 
- & 50 & 50 & 44 & 31 \\\hline
\end{tabular}
\end{table}


\comment \label{A:second:opinion} Using a second LLM to determine the correctness of the solution generated by the initial LLM significantly weakens the methodology. Since LLMs frequently hallucinate, make mistakes, and the second LLM is likely to suffer from the same memorization issues that the first LLM is being tested for I would not recommend using an LLM for this specific part of the methodology. 

\reply We agree completely: a second LLM cannot give a bill of health! Yet, several researchers are using it \cite{jensen2024software,hou2024largesurvey}. The whole purpose of this analysis is precisely to empirically test how biased is that idea, i.e. to study another possible effect of our overarching hypothesis. We have revised the paper to make this clear. We use a separate manual analysis for the ground truth (see \ref{A:manual:analysis}).

Essentially, if ``a substantial portion of the
LLM’s vulnerability repair success may be attributable
to training data memorization", we expect that an LLM generating patches will replicate the developer-generated fix seen in the training phase, and an LLM evaluating patches will easily recognize all the ones matching known developer fixes.
Thus, we expect that the LLM performance in evaluating correct patches (that possibly are developer-generated fixes leaked in the training data) will be higher than the performance in distinguishing wrong patches (never seen before, unless the patch is identical to the vulnerable version).

\comment \label{A:manual:analysis} Considering that the dataset contains only 50-100 manually curated vulnerabilities, I would highly suggest to the authors to conduct either manual analysis or find a more robust automated way to establish the correctness of the patch instead of manually sampling a few patches that passed the testing.

\reply We agree: we already proposed to use manaual analysis in RQ3. We refined it by adding the analysis of justification of the LLM, to account for a very recent result at FSE \cite{risse2025top}. This is shown in the example in Figure \ref{fig:motivating:example} (see \ref{B:example}).

To determine a minimum sample size of patches we need to manually review to understand the proportion which is correct, we use Cochran's formula which approximate the sample for a proportion: 
$$n=z_\alpha^2\cdot Pr_C(1-Pr_C)/\epsilon^2$$.

It expresses the sample size as a function of the (a priori) probability of a patch to be correct $(Pr_C)$ and the  confidence interval $(\alpha =5\%)$, and the margin of error $(\epsilon=10\%)$. The maximum uncertainty (and thus the largest sample size) is when $Pr_C=1-Pr_C=0.5$ which  yields $n=96$ samples. This must be repeated for each LLM and each plausible patch of the RQ1-2. So it is a not negligible number.

\reviewer 

\comment \label{B:research:questions} The importance of the research question(s): This RR is particularly interesting, since it wants to shed light on the real abilities of LLMs by analyzing how less visible factors (i.e., the three invisible hands) may impact the automated LLM results, in the specific context of Automated Vulnerability Repair (AVR). The paper is mainly driven by the reasoning that we all claim the good/bad results of LLMs, while not taking into account how much LLMs' results are influenced by results previously authored by humans.

The research questions are then formalized accordingly, by considering the LLM ability when errors are
discarded or introduced. Figure 2 well describes the type of experiments that will be conducted.

$\Rightarrow$ The logic, rationale, and plausibility of the proposed hypotheses: The RR considers two type of hypothesis: the one expecting statistically significant differences, the one expecting statistically equivalent differences. Both hypothesis are plausible.

\reply We appreciate that the nature of the hypothesis is clear. We further have clarified the rationale (\ref{A:elaborate:on:equivalence}) and clarified that all hypothesis are alternative hypotheses (\ref{X:TOST}, \ref{C:null:and:alternative}).

\comment \label{B:research:method} $\Rightarrow$ The soundness and feasibility of the methodology and analysis pipeline (including statistical power analysis where appropriate).
The research method is presented in detail.
Section V presents the artifacts to be considered: the dataset, the prompts, and the models.
Section VI presents the execution plan, comprising six steps.
Both sections are very informative.

\reply We appreciate the information and we have further clarified the selection of the Models

\comment \label{B:background} In the paper to be submitted, I would recommend the authors to:
provide a more accessible read to non AVR experts. A background section could be added to make the context clearer.
\reply We have rewritten the terminology and background section (see page \pageref{back:B:background}).

\comment \label{B:example} - provide a running example.
\reply We have run an example prompting ChatGPT 3.5 to fix the vulnerable function in Figure \ref{fig:motivating:example}. We propted the model to fix the vulnerability in the function at the line marked at the end with ``\textbackslash \textbackslash vulnerable line".
The prompts were presented in two different chats. The model proposed the same fix in both cases, adding a type check to line 6, Due to limited space, we only mention the result in the caption of Figure \ref{fig:motivating:example}.

\comment - Figure 2: on its bottom, it refers to RQ0 and RQ3. Please explain or remove. \label{B:exec-plan}
\reply Thank you for the suggestion, we updated Figure \ref{fig:exec-plan}. We kept RQ1 - Prompt dataset and RQ3 - Manual analysis as transversal questions, that applies to all configurations.

\comment \label{B:APR} - Section II: you refer to APR, without the acronym definition (I expect it to be Automated Program Repair, but better to clarify).
\reply Thank you for the suggestion, we now added the acronym definition on page \pageref{back:B:APR}.

\comment \label{B:prompt:ids} - Page 5 refers to P1 and P2. Since readers may confuse the IDs presented in Table III with those IDs, I would suggest reporting explicitly Prompt type 1 and Prompt type 2, or alike.
\reply We adopted the information. The prompts IDs are reported in Table \ref{tab:prompts:ours} and on page \pageref{back:B:prompt:ids} we kept L0 and L1 to enumerate how the localization information is added to the prompts, to be consistent with D1-D5 as criteria for the Dataset selection and M1-M3 as criteria for Models choice.

\comment \label{B:figure4} - Figure 4 is not explicitly cited in the paper.
\reply Thank you for the suggestion, we now explicitly cited Figure \ref{fig:helmert2} on page \pageref{back:B:figure4}.

\reviewer 

\comment \label{C:summary} Summary
The authors perform a study on Automated Vulnerability Repair (AVR) using Large Language Models (LLMs) to investigate whether previously reported successful results were influenced by hidden factors such as memorization or perfect fault localization. They replicate existing AVR experiments while intentionally introducing localization errors in the prompts provided to LLMs. The authors assess whether these introduced localization errors significantly affect patch correctness by evaluating the LLM-generated patches on two benchmarks (Vul4J and VJTrans) through regression testing, vulnerability proof tests, and manual audits. This approach allows them to quantify LLMs' effectiveness and generalization capabilities in AVR scenarios.

Strenghts
++ The topic is timely and relevant
++ The sample is sufficient for the proposed analysis

Weakness
-- Unclear methodological choices
-- Unclear hypothesis
-- Unclear impact

The authors present an interesting registered report on AVR via LLMs

\reply We are glad the reviewer appreciated the idea and we hope we clarified all doubts with this response letter.

\comment \label{C:overarching:hypothesis} 
However, since the beginning, i.e., the overarching hypothesis, presents issues. For instance, I liked the OH, but the authors then failed to properly define how they would address such a hypothesis in practice. 


\reply See also answer to comment \ref{A:elaborate:on:equivalence}. We clarified the approach as an example of Bayes theorem (see page \pageref{back:C:overarching:hypothesis}). We formulated each of our research questions to inquire into different potential manifestations of the overarching hypothesis.
For each question, we formulate hypotheses based on the effect we would expect if our overarching hypothesis were true.

\comment \label{C:polluted:training} How do they ensure the examples were not part of the training dataset? Even though the authors could create ad-hoc examples, what is the probability that some user didn't provide a very similar code excerpt online and ended up in the training pipeline of the LLM?  (First major Issue).

\reply This is really the purpose of our experiment. If the LLM just memorized code then RQ1err is true. 
To strengthen the hypothesis we have already included in the original criteria D5 the modification of the vulnerable code which is why we ended up using VJBench as they have a script to do those changes. Renaming identifiers yielded the biggest drop in performance in VJBench \cite{wu2023effective}. The third criteria for model selection (see \ref{C:LLM:selection}) also addresses this concern directly: models deployed before the benchmark's publication.

To further strengthen the hypothesis by adding an even more unlikely event (see \ref{A:elaborate:on:equivalence}) we have modified the paper by including also a new version of renamed identifiers in a \emph{language that is not English} (see page \pageref{back:C:polluted:training}). They have not appeared in the English repository of VJBench. The LLM might still be able to transform them, so the probability of a cheater LLM to map the memorized text into the new text is not zero but highly unlikely. 


\comment \label{C:LLM:selection}
These issues directly connect to the claims in the threats to validity sections: How will the author select the LLMs? The selection criteria are very vague (almost non-existent).

\reply We have rewritten the text on the selection to make it clearer. The following choices seem to be the most natual ones to us. While the other criteria are in AND, the models criteria are in OR.
\begin{description}
    \item[M1] Model evaluated in the previous pipeline
    \item[M2] Best proprietary model and best open source model in SWE LLM benchmark
    \item[M3] Same model of M1 and M2 but released before VJBench has been released (addresses also \ref{C:polluted:training}), if they are available.
\end{description}

\comment \label{C:threats} In the same vein, the threats to validity section is not discussed: the authors present specific clarification on model selection, prompt construction, model exposure, sample size, language, and ecosystem. However, the authors fail to discuss the usual threats to validity (see Wohlin et al. 2012). This is the second major issue.

\reply Each of our threats corresponds to the most relevant threats of our experiment in each of the categories in Wholin et al. (2012). The book has four categories of threats (page 102) and we have six threats covering them (Language and ecosystem, Sample size, Model exposure, LLM selection, Prompt construction, Manual evaluation) as we illustrate in Table \ref{tab::wohlin},

\begin{table}[h!]
    \caption{Mapping of our Threats to (Wohlin et al., 2012) categories. }
    \label{tab::wohlin}
    \centering
    \begin{tabular}{lp{0.45\linewidth}p{0.35\linewidth}}
   \hline
    Category
         & Definition by Wohlin et al 2012 page 102. The order of categories is from the book & Our Threats \\
    \hline
    Construct validity & To what extent the operational
measures that are studied really represent what the researcher has in mind and
what is investigated according to the research questions &
\emph{LLM selection}, \emph{Sample size} 
\\
Internal validity & 
When causal relations are
examined. When the researcher is investigating whether one factor affects an
investigated factor, there is a risk that the investigated factor is also affected by a third factor. 
& \emph{Model exposure}, \emph{LLM selection} 
\\
External validity 
& To what extent it is possible to generalize the findings, and to what extent the findings are of interest to other people outside the investigated case 
& \emph{Languages and ecosystems},  \emph{LLM selection}  \\
Reliability 
& The extent to which the data and the analysis are dependent on the specific researchers. 
& \emph{Prompt Construction}, \emph{Manual evaluation} (Section\ref{sec:artifacts} covers the other artifacts) 
\\
    \hline
    \end{tabular}
\end{table}

\comment \label{C:null:and:alternative} Moreover, the hypotheses are presented once as null hypotheses and the other time as alternative hypotheses. However, it is very unclear and confusing which alternative hypotheses are connected to which null hypothesis.

\reply  We believe this is a misunderstanding due to the use of a test that is not well known to software engineering reviewers. Please see the explanation in \ref{X:TOST}. We have made explicit with the index $H_{alt}$ that all stated hypotheses are alternative hypotheses. See also comment \ref{A:elaborate:on:equivalence}.

\comment \label{C:multiple:tests} Furthermore, both the hypotheses and RQs test multiple things simultaneously. Hence, the hypothesis testing and answering the RQs are hard to verify, measure, and address.

\reply We have clarified that each RQ tests a different manifestation of the overarching hypothesis. See also comment \ref{A:elaborate:on:equivalence}.

\comment \label{C:test:assumptions} Additionally, the authors refer to specific tests, but fail to discuss the requirements for such tests to be applied in the first place. Without accounting for normality testing, we must assume that the authors think the prerequisite of such tests holds. Still, without clear evidence and a description of the tests, it is challenging to assess the authors’ works correctly.

\reply Again we believe this is a misunderstanding similar to \ref{C:null:and:alternative}. We have not proposed new tests whose assumptions are not commonly known. See \ref{X:TOST} for what is a TOST which uses the Mann-Withney-U (MWU) as the underlying directional test. MWU is ubiquitous in Software Engineering as it makes no assumptions (in particular about normality). We believe we don't need to specify it.  Helmert Contrast is also not a test, but a way to combine tests that does not requires to use a Bonferroni correction by comparing each of the levels with the k-1 previous ones \cite{spssstatistics} (see Figure \ref{fig:helmert2}). We have added a citation on page \pageref{back:C:test:assumptions} to help the reader to find it.

\comment \label{C:RQ:contextualized} Finally, RQs are not rationalized or motivated. Considering the issues regarding hypothesis testing and the lack of RQs contextualization makes it hard for a reader, hence a reviewer, to measure the impact of the registered reports. 

\reply We have clarified this in the introduction. Please see response to \ref{A:elaborate:on:equivalence} for the contextualization and \ref{C:null:and:alternative} for the hypothesis.

\comment \label{C:Conclusions} Futhermore, the authros fail to discuss the conclusions of the Regsitered Report in which I would expect them to asses the practical relevance and the impact of their research.
 
\reply We believe we cannot make a Conclusion section in a Registered Report \emph{before} running the experiment as different results can lead to very different impacts.

\comment \label{C:final} In the current state, the registered report presents too few details and many methodological issues that make it challenging to accept.

\reply We hope we clarified all mentioned issues.

\newpage
    \pagestyle{plain}
    \setcounter{page}{1}
    \renewcommand{\thepage}{\arabic{page}}
    \setcounter{table}{0}
    \renewcommand{\thetable}{\Roman{table}}
\twocolumn
\fi

\newcommand{\gap}[1]{ %
	\medskip %
	\noindent\fcolorbox{black}{white}{%
		\parbox{0.985\linewidth}{%
			\textbf{Gap.} #1 %
		}%
	}%
	\medskip %
}%

\pagestyle{plain}

\title{Repairing vulnerabilities without invisible hands. \\ A differentiated replication study on LLMs}

\iftrue
\author{\IEEEauthorblockN{Maria Camporese}
\IEEEauthorblockA{
\textit{University of Trento}, IT \\
maria.camporese@unitn.it}
\and
\IEEEauthorblockN{Fabio Massacci}
\IEEEauthorblockA{\textit{University of Trento}, IT}
\textit{Vrije Universiteit Amsterdam}, NL\\
fabio.massacci@ieee.org}
\else
\author{\IEEEauthorblockN{Alice}
\IEEEauthorblockA{
\textit{Affiliation}, Country \\
email}
\and
\IEEEauthorblockN{Bob}
\IEEEauthorblockA{
\textit{Another Affiliation}, Another Country \\
another email}}
\fi

\maketitle

\begin{abstract}
[\emph{Background}:]  Automated Vulnerability Repair (AVR) is a rapidly emerging subfield of program repair. Large Language Models (LLMs) have recently shown promise in this area, delivering compelling results beyond traditional code generation and fault detection tasks. 
[\emph{Hypothesis}:] These results, however, may be influenced by ``invisible hands"— hidden factors such as code leakage or perfect fault localization, which allow the LLM to reproduce fixes previously authored by humans for the same code fragments.
\noindent
[\emph{Objective}:] We aim to replicate prior AVR studies using LLMs for patch generation under controlled conditions, where we deliberately introduce errors into the vulnerability localization presented in the prompts. If LLMs are merely reproducing memorized fixes, both small and large localization errors should result in a statistically equivalent number of correct patches, as each type of error should steer the model away from the original fix.
\noindent
[\emph{Method}:] We introduce a pipeline for repairing vulnerabilities in the Vul4J and VJTrans benchmarks. The pipeline intentionally offsets the fault localization by n lines from the ground truth. An initial LLM generates a patch, which is then reviewed by a second LLM. The resulting patch is evaluated using regression testing and vulnerability proof tests. Finally, we conduct a manual audit of a sample of patches to assess correctness and compute the statistical error rate using the Agresti-Coull-Wilson method.

\end{abstract}


\section{Introduction}
\label{sec:introduction}
Large language models (LLMs) are increasingly proposed for automated vulnerability repairs (AVR) \cite{zhou2024large}. At the time of submission, Google Scholar reports 483 works on the topic from 2024, excluding ArXiv papers, SoKs and systematic literature reviews.
LLMs results (e.g., \cite{huang2025comprehensive}) seem more promising than traditional repair approaches based on testing (e.g., \cite{bui2024apr4vul}).
APR4Vul tested traditional program repair methods on the Vul4J dataset \cite{bui2022vul4j} and obtained at best 5 correct patches. LLM methods obtained much better results: Codex was able to fix 10 vulnerabilities without fine-tuning \cite{wu2023effective}, later works provided 12 perfect fixes \cite{sagodi2024reality}, and up to 14 patches \cite{kulsum2024case}.

Yet, replicating LLMs' success is challenging \cite{zhou2024large}, mirroring issues previously observed with deep learning models \cite{sejfia2024toward}. We argue that three ``\emph{invisible hands}'' could be at play: \emph{code leakage, completion vs patching, and perfect localization}. 
These factors could help LLMs achieve ``exceptional'' results that testing-based program repair methods did not achieve. 

The first ``hand" involves code leakage: as a benchmark for vulnerability repair is eventually published, the new LLMs are likely to be trained on that very dataset, and the LLMs used for the original study might no longer be available. For instance, replicating the studies in \cite{wu2023effective, pearce2023examining} require access to Codex, which is now deprecated. Further, suppose we wanted to use PrimeVul \cite{ding2024vulnerability}, a high-quality dataset with robust, albeit automatically verified, vulnerable and fixed functions. 
The corresponding paper was published at ICSE in May 2025, and therefore, GPT-4o-mini, deployed in May 2024, would seem to be a good candidate (it is used in 34 of the mentioned LLM papers). 
Unfortunately, the dataset was published on GitHub in March 2024 and includes commits pushed even earlier, so the possibility of data leakage cannot be ruled out.

The second and third invisible hands stem from perfect fault localization and prompt design. 
Providing the exact location of the vulnerability is often the gold standard in program repair, e.g., in \cite{bui2024apr4vul} (non-LLM) and \cite{li2024hybrid} (LLM-based). 
By itself, this is a reasonable approach to distinguish success in vulnerability localization from success in patch generation. 
The key problem happens when it is combined with prompt designs that frame repairs as code completion: the fragment to be patched is replaced with a placeholder, and the LLM is asked to complete the code (e.g. \cite{pearce2023examining, kulsum2024case, wu2023effective}). Given the likely leakage of code (see first invisible hand), LLMs may simply reproduce the correct fix via completion, a task where LLMs excel \cite{zhang2024instruct}. 

\finding{\textbf{\textit{Overarching Hypothesis.}} A substantial portion of the LLM's vulnerability repair success may be attributable to training data memorization, facilitated by precise fault localization.}
 
Our is a \emph{skeptical replication study}, aimed at disentangling the potential influence of ``invisible hands" on the success of LLM-based AVR. How can we remove the influence of these hands in order to test the hypothesis? 
\fixed{\reviewnote{A:elaborate:on:equivalence}We need to measure some events that are very unlikely to happen if the LLM is generalizing (``ideal" LLM) vs the LLM is just relying on memorization (``cheater"
LLM). Bayes theorem would then allows us to distinguish the posterior probability. Some preliminary evidence in this respect has been already explored for normal bug fixing \cite{kong2025demystifying}.}
Our idea is to \emph{systematically introduce errors} in the localization of vulnerabilities.

\fixed{If the LLM is ideal and genuinely generate repairs based on the localized vulnerability, then it is very unlikely that fixing a line that is completely off will generate a correct fix. \reviewnote{A:elaborate:on:equivalence:second}\reviewnote{C:overarching:hypothesis}
And even less the fix of the developer...
Conversely, if the LLM only memorizes fixes during training, then it is very likely that it will ignore the localization suggested by the prompt and propose the already known fix (if at all).
Either way, we expect the LLM to reach the same results irrespective of the magnitude of the displacement offset (it's cheating).}

To strengthen the result, we add a second step: we \emph{ask a second LLM to review the result of the first LLM before testing}. \fixed{This is considered a good solution to scale \cite{hou2024largesurvey}} albeit a recent theoretical analysis \cite{camporese2025using} has shown that MLs patch-reviewer model can only be useful if precision and recall are above a certain threshold. \fixed{Worse, if our overarching hypothsis is true the second LL M is likely to suffer from the same memorization issues that the first LLM is being tested. If LLMs merely reproduce and recognize fixes they have already seen, we expect cheater LLMs as reviewers to recognize (memorized) plausible patches better than other plausible but wrong patches. 
As soon as we perturb the pipeline so that recognition is hampered, such difference will disappear. Localization rrors will not make a difference as they are ignored by cheaters.}

Operationally, we build on the pipelines from \cite{wu2023effective} and \cite{kulsum2024case}: we assess the impact of different prompt designs and \fixed{include in our input vulnerabilities obtained through code transformations so that they have no corresponding published fix. This contribute to isolate model generalization from memorization.} We will validate generated patches using both regression tests and Proof of Vulnerability (PoV) tests\footnote{A proof of vulnerability in test-based vulnerability repair \cite{pinconschi2021comparative,bui2024apr4vul, zhang2022program} is a test which flips between the vulnerable and non-vulnerable code fragments.}

\section{Terminology and background}
\label{sec:terminology}

\fixed{\reviewnote{B:background} Automated Program Repair (APR) \reviewnote{B:APR} aims to support developers by automatically fixing bugs in software. End-to-end APR pipelines cover the full repair process: locating the faulty code (fault localization), generating a patch, and validating it to ensure correctness \cite{shen2020survey}. The ultimate goal is to produce patches that can be reviewed and accepted by developers.

To evaluate APR tools, researchers use benchmarks composed of buggy programs and associated test suites. A patch is considered \textit{plausible} if it passes all available tests automatically. However, only \textit{correct} patches are semantically equivalent to the original developer fix, while \textit{overfitted} patches pass tests but fail to fix the root cause of the bug \cite{pinconschi2021comparative, bui2024apr4vul, wu2023effective}.

Automated Vulnerability Repair (AVR) is a specialization of APR focused on fixing security-related bugs—vulnerabilities that attackers could exploit. In addition to preserving functionality, AVR patches must restore or maintain security properties. As a result, AVR benchmarks often include not only unit tests but also \textit{Proof of Vulnerability (PoV)} tests \cite{bui2022vul4j}, which demonstrate that the original bug can be exploited.

In the literature, AVR-generated patches that pass all tests are also referred to as \textit{plausible}, as in general APR \cite{wu2023effective}, or as \textit{End-to-End (E2E) tested patches} \cite{bui2024apr4vul}. Upon manual inspection, such patches may turn out to be \textit{correct}, \textit{overfitted}, or \textit{security-fixing}—patches that remove the vulnerability but silently introduce regressions that break functionality \cite{bui2024apr4vul}.}


\section{Related works}
\label{sec:rw}

\subsubsection{Evaluation pipelines for AVR}
\label{subsec:evaluations:avr}
Evaluating tools for repairing vulnerabilities required the development of benchmarks provided with functional and security tests. Pinconschi et al. \cite{pinconschi2021comparative} developed \textsc{SecureThemAll} to test APR techniques for repairing 55 C/C++ security faults from the DARPA Challenge Sets \cite{CGC}, which approximate real-world vulnerabilities and include functional and PoV tests. The 10 APR tools tested received only C source code and its test suite. Bui et al. \cite{bui2024apr4vul} proposed APR4Vul to evaluate general APR techniques on vulnerabilities in Vul4J \cite{bui2022vul4j}, a manually curated benchmark of 79 reproducible Java vulnerabilities, each with functional and PoV tests. APR4Vul tools were provided with exact, manually defined vulnerability locations to isolate the patch generation task from localization errors.

\subsubsection{Evaluations of LLMs for AVR}
\label{subsec:llms:avr}
\begin{figure*}[h]
    \centering
    \includegraphics[width=\textwidth]{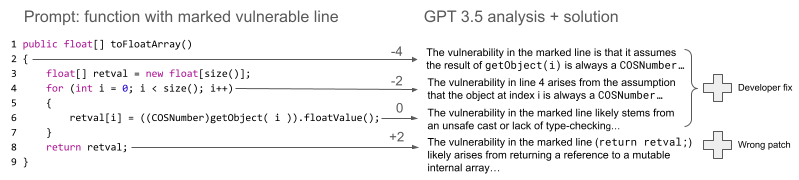}
    \vspace{-7mm}
    \caption{While different studies tested LLMs for repair starting from the exact vulnerability localization, we investigate the impact of errors in the localization for different approaches. For example, in the approach proposed by Wu et al. \cite{wu2023effective}, the LLM is prompted to substitute exactly the vulnerable lines, so even small displacements would prove disruptive. \fixed{Here we prompted GPT3.5 (used by Kulsum et al. \cite{kulsum2024case}) to fix the vulnerable function, but gave in the prompt wrong information about the vulnerable line. In the figure, we have two negative and one positive offsets and the corresponding responses. When the offset is 4 lines above, pointing to just a curly bracket (line 2), the model still generates the developer fix. How is this likely to happen in the absence of memorization?}}
    \label{fig:motivating:example}
\end{figure*}

Different studies proposed testing LLM to repair software vulnerabilities. Pearce et al. \cite{pearce2023examining} tested proprietary and open-source LLMs, including a locally trained model, for zero-shot vulnerability repair in C/C++. They evaluated these against synthetic vulnerabilities and 12 real-world CVEs using six prompt templates (Table~\ref{tab:prompts:codexvr}) to present the exact vulnerability location. Notably, prompts like ``n.h.'', ``s.1'', and ``s.2'' omit the faulty code, framing the task as code generation rather than repair—contrary to AVR's assumption that the defect is known.
\begin{table}[t]
    \caption{Prompt templates used by Pearce et al.~\cite{pearce2023examining}.}
    \label{tab:prompts:codexvr}
    \begin{tabular}{lp{0.85\linewidth}}
    \toprule
    ID & Description \\ \hline
    n.h.      & No Help - deletes the vulnerable code/function body and provides no additional context for regeneration.                                                                                                                  \\ \hline
    s.1       & Simple 1 - deletes the vulnerable code/function body and adds a comment ‘bugfix: fixed {[}error name{]}’.                                                                                                                 \\ \hline
    s.2       & Simple 2 - deletes the vulnerable code/function body and adds a comment ‘fixed {[}error name{]} bug’.                                                                                                                     \\ \hline
    c.        & Commented Code - After a comment ‘BUG: {[}error name{]}’, it includes a ‘commented-out’ version of the vulnerable code/function body followed by the comment ‘FIXED:’. After this it appends the first token of the original vulnerable function. \\ \hline
    c.a.      & Commented Code (alternative) - same as c. , but commented in the alternative style for C /* and */ rather than //                                                                                                                        \\ \hline
    c.n.      & Commented Code (alternative), no token - same as c.a., but with no ‘first token’ from vulnerable code.                                                                                                                    \\ \bottomrule
    \end{tabular}
\end{table}
Wu et al.~\cite{wu2023effective} proposed a framework to test DL-based tools for AVR in Java.
They extended the Vul4J benchmark \cite{bui2022vul4j} and proposed VJBench-trans, a new dataset of transformed versions of the collected single-hunk vulnerabilities.
The transformation should mitigate the advantage for LLMs that were already exposed to the testing data during their training.
The prompt templates that the authors use are reported in Table~\ref{tab:prompts:effective}, and all of them but ``Codex'' present a code completion rather than a vulnerability repair problem.
\begin{table}[t]
    \footnotesize
    \caption{Prompt templates used by Wu et al.~\cite{wu2023effective}.}
    \label{tab:prompts:effective}
    \begin{tabular}{l@{}p{0.8\linewidth}}
    
    \toprule
    Model & Input Format \\ 
    \midrule
    Codex & Comment  buggy lines (BL) with hint ``BUG:'' and ``FIXED:'' \\ %
    & Prefix prompt: Beginning of the buggy function to  BL comment \\
    & Suffix prompt: Line after BL comment to end of the buggy function\\
    \midrule
    CodeT5      & Mask  buggy lines with $<$extra\_id\_0$>$ and  input the  buggy function \\
    \midrule
    CodeGen     & Input beginning of the buggy method to line before buggy lines \\
    \midrule
    PLBART      & Mask  buggy lines with $<$mask$>$ and  input the  buggy function \\
    \midrule
    InCoder     & Mask  buggy lines with $<$mask$>$ and  input the  buggy function \\
    
    \bottomrule
    \end{tabular}
\end{table}
Since they only prompt LLMs to substitute the lines they mark as vulnerable, their approach heavily rely on the exact vulnerability localization,
Even a slight displacement in line localization, as in Figure \ref{fig:motivating:example}\fixed{\reviewnote{B:example}}, prevents the LLM from generating a correct repair.
Kulsum et al. \cite{kulsum2024case} replicated the work by Pearce et al. with GPT‑3.5~\cite{openai2023gpt35} and added chain-of-thought prompts and iterative feedback by external tools to improve the repair process.
They tested their approach against 60 real-world CVEs (10 in C and 50 in Java from the Vul4J~\cite{bui2022vul4j} and VJBench~\cite{wu2023effective} benchmarks).
Their first prompt follow the same formats proposed by Pearce et al.

Fu et al. \cite{fu2024aibughunter} proposed \textsc{AIBugHunter}, an ML-based software vulnerability analysis tool for C/C++ for Visual Studio Code. 
They integrate their tools LineVul \cite{fu2022linevul} for vulnerability localization and VulRepair \cite{fu2022vulrepair} for vulnerability repair. However, the pipeline is only evaluated via a qualitative user study, making its objective performance hard to assess.

\gap{None of the current evaluations of LLM-based approaches perform a measurable evaluation when the vulnerability localization provided to the models is not perfect.}
\vspace{-1mm}

\subsubsection{LLM second opinion}
As code reviewing is a tedious and expensive process for software development, different studies have proposed leveraging LLMs to ease it.
Jensen et al. \cite{jensen2024software} investigated using LLMs in reviews to evaluate both code security and functionality. 
The performance of proprietary models in particular is impressive (over 95\% accuracy for vulnerability detection, over 88\% F1 score for functionality validation), however it is not clear to what extent this could be due to data leakage.
Recent works from Jaoua et al. \cite{jaoua2025combining} and Kavian et al. \cite{kavian2024llm} combine the use of static analyzers and LLMs to review and improve the quality of developer and LLM-generated patches, respectively.


\section{Research Statement}
\label{sec:res_statement}
\begin{figure*}[h]
    \vspace{-3mm}
    \centering    \includegraphics[width=\textwidth]{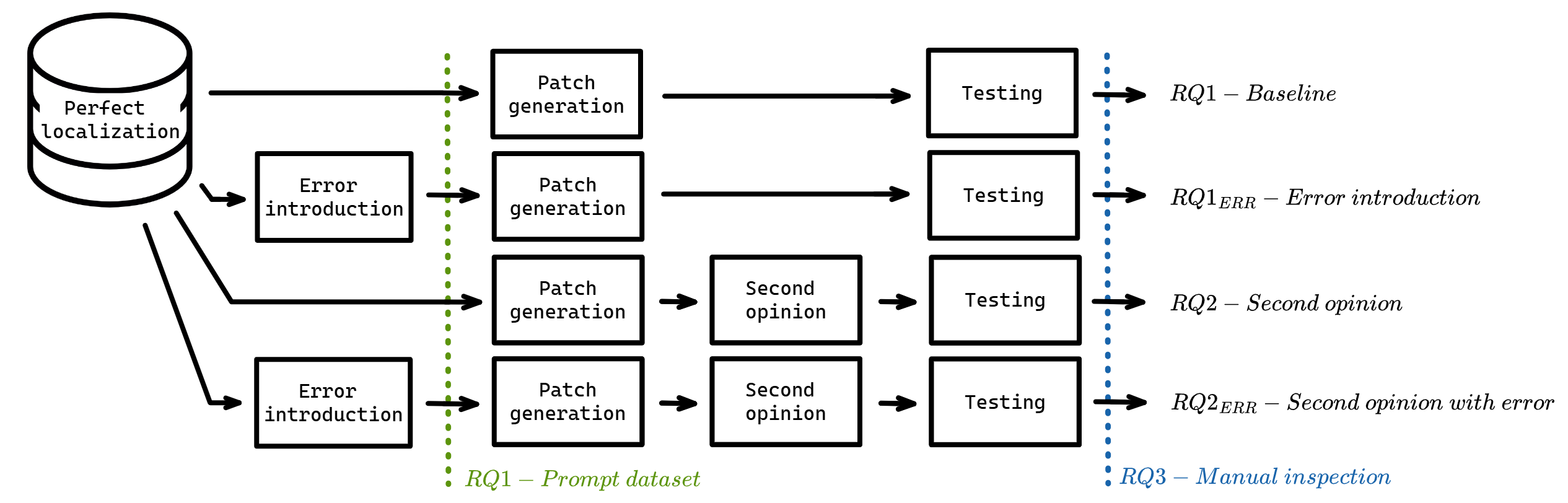}
    \vspace{-8mm}
    \caption{Execution plan.}
    \label{fig:exec-plan}
    \vspace{-5mm}
\end{figure*}

Figure~\ref{fig:exec-plan} summarizes our research questions, which start from the differentiated replication of the setups in ``How Effective Are Neural Networks for Fixing Security Vulnerabilities" \cite{wu2023effective} and VRPilot~\cite{kulsum2024case}.
Building on the mentioned studies, we use different prompt templates to investigate how they affect the repair capabilities of LLMs.

\begin{enumerate}
    \item[]\hspace*{-4ex}$\mathbf{RQ1}$\textbf{ - Reproducibility baseline.} \textit{How many vulnerabilities can LLMs fix when provided with their exact localization and how sensitive to the prompt are the results?}
\end{enumerate}

\noindent If our overarching hypothesis is true, the only way for a ``cheater'' LLM to produce a correct patch is when it was exposed to the developer-generated fix during training. 
\fixed{Then providing additional information as the vulnerability type or its line-level localization, will neither help nor confuse the LLM in recognizing the previously seen vulnerability more than the code itself, and thus it will not affect the repair results.

\reviewnote{X:TOST}To scientifically validate this hypothesis we cannot use the traditional approach used in software engineering which tests for significant differences. We claim that two treatments (showing the line or providing vulnerability information) have the same effect. This is the same idea behind statistically proving that a generic label drug has the same effect of a branded label drug \cite{food2001guidance}. We need to use a different approach, and namely prove \emph{significant equivalence} \cite{meyners2012equivalence} and not just no significant difference. In this set-up our alternative hypothesis are hypothesis about equivalence.}

$H^{alt}_{\mbox{equiv-info}}$: \textit{There is statistically significant equivalence in performance between patches generated with prompts that provide information on the type of vulnerability and prompts that only mention the presence of a security defect.}

\fixed{Since LLMs are good at translating we expect the hypothesis to hold for the transformed vulnerabilities in which identifiers are in a language different than English. How likely is an LLM that is not memorizing to generate a patch that is identical to the known patch but for renamed identifiers?} 

$H^{alt}_{\mbox{equiv-loc}}$: \textit{There is statistically significant equivalence in performance between patches generated with vulnerability localization prompted at function and  line level.}

\noindent  Then, we introduce a controlled error in the vulnerability localization at the line level and repeat the evaluation. 
\begin{enumerate}
    \item[]\hspace*{-4ex}$\mathbf{RQ1_{ERR}}$\textbf{(Error introduction)} \textit{How many vulnerabilities can LLMs fix when line-level localization is shifted?}
\end{enumerate}

\noindent Prompting an LLM with an incorrect vulnerability localization may lead to two possible outcomes: the model might ignore the incorrect line and regenerate the known fix, or it might fail to recognize the original fixing pattern and produce an incorrect patch — even in cases where it previously succeeded with accurate localization. If the second outcome plays a significant role, we would expect to see a decline in performance when localization errors are introduced.
We hypothesize that any deviation from the correct line will be enough to trigger this effect, and therefore we do not expect meaningful performance differences between small ($\le2$) and larger \fixed{\reviewnote{A:displacements} ($4$, $8$) displacements.
From 0 and 1 the difference will not be enough because we cannot distinguish the ``honest" and the ``cheating" because of the git diff window.}

$H^{alt}_{\mbox{equiv-err-order}}$: \textit{There is statistically significant equivalence in the performance between patches generated with prompts for which the line-level location is shifted by a small ($<3$ in magnitude) and a considerable ($>3$) offset.}

\fixed{When the LLM is given a negative offset (\emph{before} the vulnerable line), the cheater LLM will just respond with the known completion. When a positive offsets (\emph{after} the known vulnerable line) is given, the LLM might just hallucinate or have a very poor performance (see Figure \ref{fig:motivating:example}). So we expect a different behavior between positive and negative offsets.} 

In the second part of the experiment, the reviwer LLM will evaluate the patches generated in the first phase by discarding wrong patches and saving the ones that seem correct.

\begin{enumerate}
    \item[]\hspace*{-4ex}$\mathbf{RQ2}$\textbf{ - Second opinion.} \textit{How many incorrect patches can LLMs detect before the testing phase?}
\end{enumerate}

\noindent \fixed{If our overarching hypothesis holds, and LLMs just succeed because they recognize the vulnerable code and repropose fixes they have seen in their training, both patch generation and reviewing are reduced to similar pattern-matching tasks.

$H^{alt}_{\mbox{diff-review}}$: \textit{There is statistically significant difference in the review performance between correct patches than wrong but still plausible patches.}

$H^{alt}_{\mbox{equiv-review}}$: \textit{There is statistically significant equivalence in the review performance between patches in the original language and patches with renamed identifiers.}}

We will also run the reviewing process for patches generated with the controlled displacement in the prompt.
\begin{enumerate}
    \item[]\hspace*{-4ex}$\mathbf{RQ2_{ERR}}$\textbf{ - Second opinion with error.} \textit{How many incorrect patches can LLMs detect before the testing phase?}
\end{enumerate}
 
Finally, we will perform manual validation of a subset of the patches surviving the testing cases in each process.
\begin{enumerate}
    \item[]\hspace*{-4ex}$\mathbf{RQ3}$\textbf{ - Manual inspection.} \textit{How many patches that pass tests are semantically correct?}
\end{enumerate}

\fixed{We will also investigate whether the justification of the LLM is also correct, following the preliminary findings of \cite{risse2025top}.}

\section{Artifacts}
\label{sec:artifacts}
    \subsubsection{Dataset}
    \label{subsec:dataset}  
    We consider the following criteria:
    \begin{trivlist}
        \item D1: \textit{The dataset was used for evaluating LLMs for AVR}, to assess the impact of imperfect vulnerability localization and cross-validation on the repair performance of the models.
        \item D2: \textit{The dataset contains real-world data}, as synthetic data could distort performances (see the difference on synthetic and real-world data by Pearce et al.~\cite{pearce2023examining}).
        \item D2: \textit{The dataset includes regression and PoV tests}, since we aim to perform an extensive evaluation of LLMs for AVR. We need a reliable, automated way to determine which patches preserve the functionality of the code and which successfully repair the target vulnerability.
        \item D3: \textit{The dataset includes diverse entries}, as different types would grant the most realistic performance results.
        \item D4: \textit{ The dataset contains single function vulnerabilities}, as addressing faults in different functions with a single LLM query would pose challenges that are currently impractical.
        \item D5: \textit{The dataset contains refactored vulnerabilities} \fixed{\reviewnote{C:polluted:training}(e.g.,control flow has been changed \cite{wu2023effective} or identifiers are renamed in a language that is not English), so that the corresponding developer fix is not available. }
    \end{trivlist}
    For the listed criteria, our choice falls on the dataset VJBench and VJBench-trans~\cite{wu2023effective} built to extend the benchmark Vul4J~\cite{bui2022vul4j}.
    These datasets contain Java vulnerabilities, of which at least 50 single-hunk vulnerabilities (Wu et al. selected them in 2023, but Vul4J was expanded since then).
    \fixed{We aim to widen this selection to single-function vulnerabilities and use the scripts of the refactoring process of VJBench-trans to generate vulnerabilities that the LLMs have never seen.}
     
    \subsubsection{Prompts}
    \label{subsec:prompts}  \fixed{\reviewnote{B:prompt:ids}Table~\ref{tab:prompts:ours}} summarizes the four prompt templates we plan to use, building on previous works that evaluated LLMs for AVR \cite{pearce2023examining, wu2023effective} with two levels of information: 
    \begin{trivlist}
        \item \fixed{L0:} \textit{The prompt contains the vulnerable lines}, as the only available localization (so P1 and P3). To generate a correct patch from the function alone, without clues, the model must have seen a very close human patch. 
        \item \fixed{L1:} \textit{The prompt includes the functionally correct vulnerable code segment, with an indication of the security flaw.} (P2, P4) Rather than commenting the line as done in ythe state of the art, by providing a functionally correct code segment, we ensure that if the application of a patch generated by the LLM does not result in the project compiling and passing the regression tests, the model is to be blamed. 
        Thus, we propose to insert the comments marking the vulnerable lines as a suffix rather than as a prefix of the code they contain.
    \end{trivlist}
    
    \begin{table}[t]
        \centering
        \caption{Prompt templates we plan to use for this study.}
        \label{tab:prompts:ours}
        \vspace{-2.5ex}
        \fixed{\begin{tabular}{lp{0.85\linewidth}}
        
        \toprule
        ID  &
             Description \\ \midrule
    
             P1 & General information + vulnerable function + output request \\
        
             P2 & P1 + vulnerable lines marked with the suffix ``// BUG'' \\ 
    
            P3 & P1 + vulnerability description \\ 
    
            P4 & P3 + vulnerable lines marked with the suffix ``// BUG''  \\ 
        
        \bottomrule
        \end{tabular}}
        \vspace{-6mm}
    \end{table}

    
    
        
    


 
    In all cases we provide the code of the vulnerable function, but in the first one we add no additional information (as if the vulnerability localization was performed by ReVeal~\cite{chakraborty2021deep}), the second provides line-level localization (like LineVul~\cite{fu2022linevul} could do), the third one provides the vulnerability type (as ChatGPT was prompted to do in \cite{fu2023chatgpt}) and the last one both line-level localization and vulnerability type (as a SAST tool could do~\cite{jasz2022end}).
    Besides the code and eventually the vulnerability type, each prompt will include a request to generate a new version of the given function to fix the present vulnerability.
    \fixed{If other prompts are equivalent to the bare one, we know they all memorize and the additional information is irrelevant.}

\subsubsection{Models}
    \label{subsec:models}
    To select the LLMs for our replication study, we adopt the following criteria:
    \begin{trivlist}
        \item M1: \textit{Models should have already been evaluated in similar pipeline.} Both the evaluations of LLMs for AVR performed by Wu et al.~\cite{wu2023effective} and Pearce et al.~\cite{pearce2023examining} respectively found that Codex~\cite{openai2021codex} was the best performing model.
        Unfortunately, Codex models are no longer available, so Kulsum et al.~\cite{kulsum2024case} had to base their approach on ChatGPT 3.5~\cite{openai2022chatgpt}, 
        We build on their choice and adopt it as the first model in our study.
        \item \fixed{M2: \textit{Best proprietary model and best open source model on LiveBench~\cite{whitelivebench} } At the time of writing, the best-performing proprietary model is GPT-o4-Mini High~\cite{openai2025o4mini} and the best open-source model is DeepSeek R1 Distill Qwen 32B~\cite{guo2025deepseekr1}.}
         \item \fixed{M3: \textit{Same model of M1 and M2 but released before VJBench has been released}. to further avoid memorization.}
\end{trivlist}

\section{Execution Plan}
\label{sec:exec_plan}


First, we outline the general pipeline setup, then we further explain each step in the later subsections.

\subsubsection{Setup}
\label{sssec:setup}
The full pipeline setup is shown in Figure~\ref{fig:exec-plan}, under the $RQ2_{ERR}-$Second opinion with error sequence. We extract prompts from the dataset built in Section~\ref{subsec:prompts}. Depending on the RQ, prompts are either sent directly to the patch-generation LLM or first undergo error injection, where line-level vulnerability localization is deliberately shifted. The LLMs are prompted to produce a fixed version of the vulnerable function that preserves original functionality. For RQs involving a second-opinion assessment, the generated functions are passed to a second LLM, which outputs \textit{TRUE}'' if the function is correct, or \textit{FALSE}'' if the vulnerability persists or functionality is broken. All generated functions are tested by replacing their vulnerable counterparts in the original project, compiling, building and running regression and PoV tests.

\subsubsection{Error introduction}
We introduce a controlled error in vulnerability localization. As in \textit{Setup}, each LLM receives the full vulnerable function as input. Since correction is impossible when localization is incorrect at the function level, we apply this step only when line-level localization is also provided. \fixed{For each such prompt, we generate six variants by shifting the marked line by 2, 4, 8 lines—either above (e.g., -2) or below (e.g., +2) the correct location.} Variants that fall outside the function's scope are discarded. If testing all prompts becomes too costly, we apply a Taguchi design~\cite{mitra2011taguchi,massacci2024addressing} to design a balanced experiment and reduce the number of evaluations.

\subsubsection{Patch generation}
We query each LLM (or LLM-based setup, when we replicate VRPilot \cite{kulsum2024case}) with all prompts from \S\ref{subsec:prompts}, or their error-injected versions. Each model is queried exactly once per prompt version and tasked with generating a repaired version of the vulnerable function. The proposed function is then extracted and passed to the validation process.

\subsubsection{Second opinion}
For the RQs that require a second LLM opinion, we prompt the second model to assess whether the patch generated by the first model is actually correct.
We provide the second model with the input used for the patch-generation phase and the new version of the function proposed, and then we instruct the model to output ``\textit{TRUE}" if the new version of the function fix the security vulnerability and maintains the code functionality or  ``\textit{FALSE}" otherwise.

\subsubsection{Testing}
Each attempt of a new patch is inserted in the software project in place of its original, vulnerable counterpart.
Then, we attempt to compile the project, and eventually we build the software and perform regression and PoV tests.

\subsubsection{Manual inspection}
To answer RQ3, we perform a manual inspection of a random sample of the patches that pass the testing phase. Two authors will independently go through each patch in the sample and assess the correctness of the patch. 
We refer to Bui et al.~\cite{bui2024apr4vul} and use ``Correct'' (fix the vulnerability and do not break the functionality), ``Partially correct'' (fix the vulnerability but break the functionality), and ``Wrong'' (do not fix the vulnerability) for the classification. 
The two reviewers will then resolve any disagreements through discussion, or possibly with a third individual as adjudicator.
\fixed{We also mark whether the LLM reasoning is right or wrong.}

\section{Analysis}

For the analysis, we count the patches for which the project still compiles when they are substituted for their vulnerable counterpart, and the patches passing regression and PoV tests.

\subsubsection{RQ1} 
\begin{figure}[t]
    \includegraphics[width=\columnwidth]{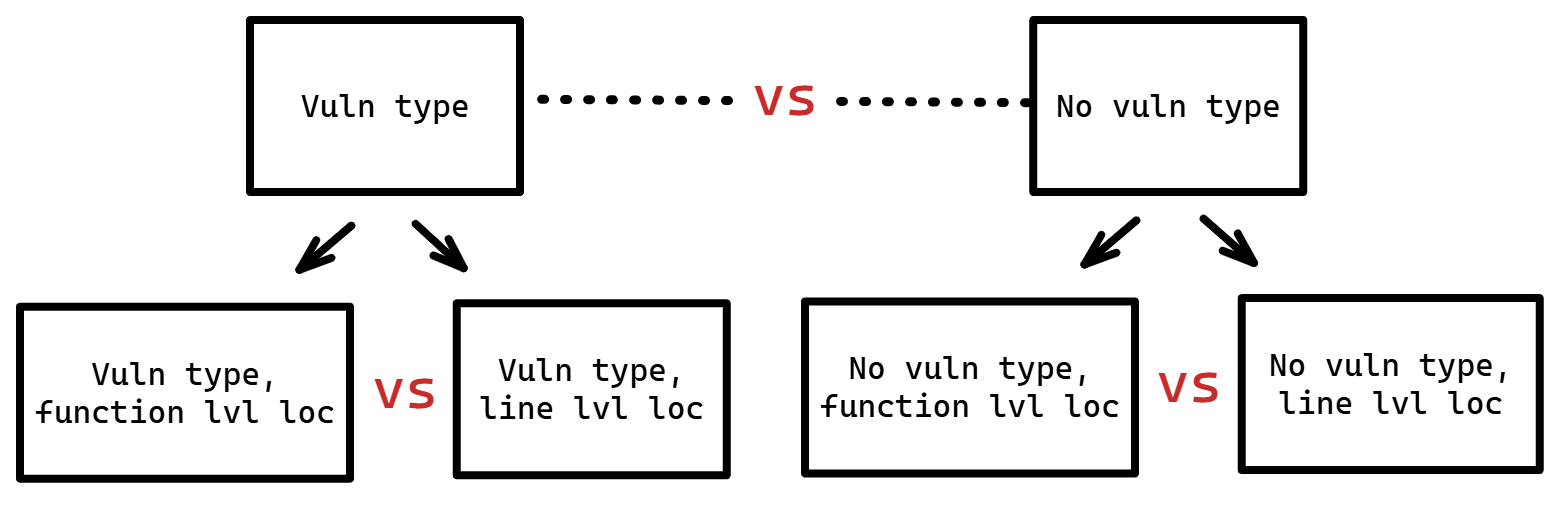}
    \caption{Helmert contrast for $RQ1$ }
    \label{fig:helmert}
\end{figure}

To verify the hypotheses about the prompt impact that we presented in Section~\ref{sec:exec_plan}, we will measure statistically significant effects on the number of produced patches passing each phase (generation, functional validation, security validation, manual inspection).
\fixed{We use \reviewnote{C:test:assumptions}\reviewnote{X:TOST} TOST (two one-sided tests) for testing equivalence \cite{schuirmann1981hypothesis,food2001guidance,meyners2012equivalence} with Mann-Whitney-U (MWU) as the underlying directional test with Helmert contrast for multiple comparisons \cite{spssstatistics}  } in the configuration shown in Figure \ref{fig:helmert}. 
We  compare the patches generated by prompts providing the type of vulnerability against those without any information on the type of defect to be repaired.
Then, we compare the patches generated with information about type and line-level localization, and the ones generated with the type and the function-level localization.
Finally, we compare the patches generated without the type, but with the line-level localization of the vulnerability, and the patches generated without the type and with the function-level localization.

\subsubsection{$RQ1_{ERR}$} For each displacement of the line-level vulnerability localization (0-baseline, 2, 4, 8), we compare the number of patches that pass each pipeline phase. As for the previous RQ, we use Mann-Whitney tests with Helmert contrast to measure the differences between different groups.
\begin{figure}[t]
    \includegraphics[width=0.7\columnwidth]{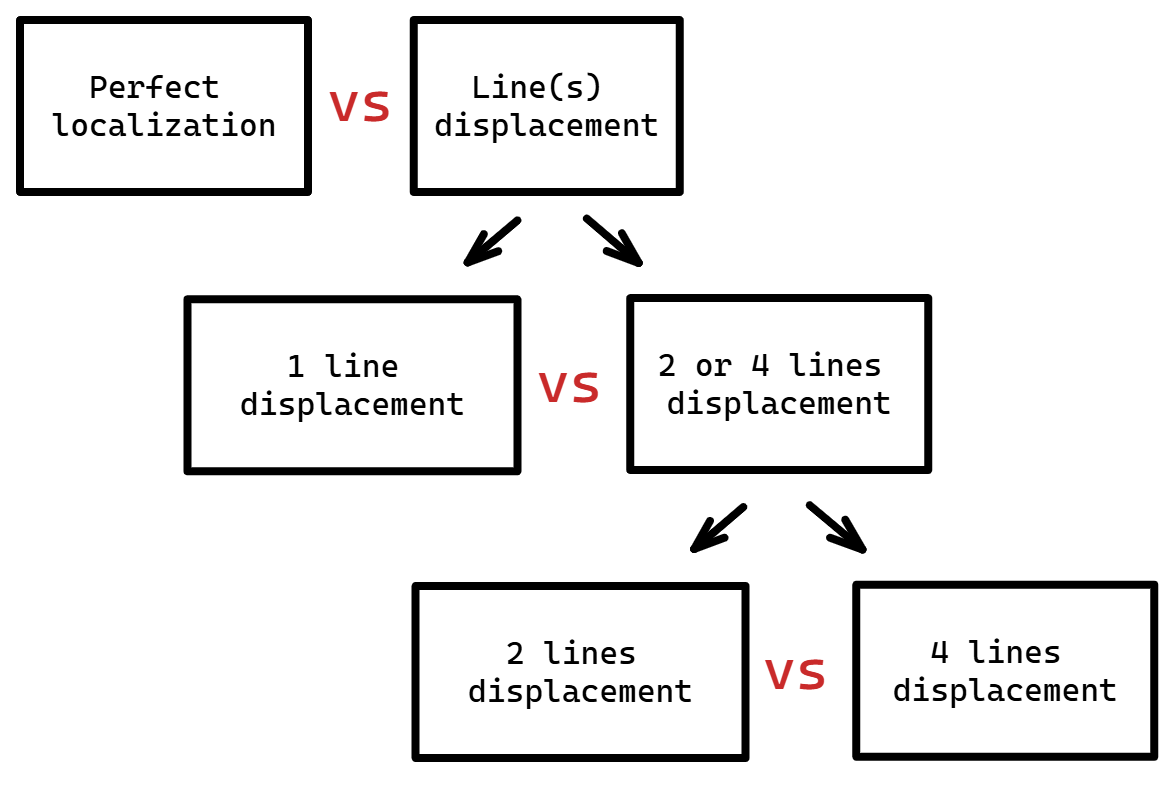}
    \caption{Helmert contrast for $RQ1_{ERR}$}
    \label{fig:helmert2}
    \vspace{-5mm}
\end{figure}
\fixed{As represented in Figure \ref{fig:helmert2},\reviewnote{B:figure4}} first, we compare the baseline with all the patches generated with errors in the line-level vulnerability localization. 
Then, the patches with the smallest error ($2$) against the ones with higher ones ($4$, $8$).
We need a TOST to show a significant equivalence, while a single MWU-will be sufficient to prove a significant difference.

\subsubsection{$RQ2$ and $RQ2_{ERR}$} We perform the same analysis as RQ1 and $RQ1_{ERR}$, but we also measure the number of patches that pass the second LLM opinion phase.

\subsubsection{$RQ3$} The Agresti-Coull-Wilson method \cite{agresti1998approximate} allows to establish
accurate and reliable confidence intervals for proportions.
This interval will enable us to extrapolate the false positive rate from sampled subsets of patches. 
\fixed{By using Cochran's formula we need to analyze 96 patches 
\reviewnote{A:manual:analysis} to have 95\% confidence interval with a 10\% margin of error \cite{spssstatistics}}
    \section{Threats to validity}
\begin{trivlist}
    \item \textit{Language and ecosystem.} Our study focuses on Java vulnerabilities. While aligned with most AVR datasets, this limits generalizability to languages like C/C++ or Python, which differ in structure and vulnerability patterns.

    \item \textit{Sample size.} Vul4J and its derivatives (e.g., VJBench, VJTransBench) contain only 50–100 manually curated, reproducible vulnerabilities with tests. While valuable for evaluation, this small size may limit statistical robustness and generalizability.

    \item \textit{Model exposure.} LLMs may have seen vulnerabilities from Vul4J or VJBench during training. This risk, especially for proprietary models, is mitigated by using refactored versions and entries from VJTransBench for which there are no published fixes available.

    \item \textit{LLM selection.} We evaluate top-performing models available at the time. As newer versions may quickly outperform them, we include both proprietary and open-source LLMs and document model versions used.

    \item \textit{Prompt construction.} Prompts are manually crafted from prior templates. Minor variations may affect LLM responses. We aim to release all prompts to support replicability.

    \item \textit{Manual evaluation.} RQ3 involves subjective judgments, particularly for partially correct patches. Two authors label independently, resolving disagreements by consensus, though ambiguity may remain.

\end{trivlist}

\section{Acknowledgements}

This work has been partly supported by the European Union (EU) under Horizon Europe grant n.\ 101120393 (Sec4AI4Sec), by the Italian Ministry of University and Research (MUR), under the P.N.R.R. – NextGenerationEU grant n.\ PE00000014 (SERICS) CUP E63C24000590001, and by the Nederlandse Organisatie voor Wetenschappelijk Onderzoek (NWO) under grant n.\ KIC1.VE01.20.004 (HEWSTI).

\section{CRediT Author Statement}
Conceptualization MC, FM;
Methodology MC, FM; 
Software 
Validation MC, FM;
Formal analysis FM, MC;
Investigation MC;
Resources MC, FM;
Data Curation MC; 
Writing - Original Draft MC;
Writing - Review \& Editing MC, FM;
Visualization MC;
Supervision FM;
Project administration FM;
Funding acquisition FM.

\clearpage

\bibliographystyle{IEEEtran}
\bibliography{short-names, biblio}

\end{document}